\documentclass[reprint,amsmath,amssymb,aps,prb,superscriptaddress]{revtex4-1}
\usepackage{natbib}
\usepackage{makeidx}
\usepackage{url}
\usepackage[colorlinks=true,linkcolor=blue]{hyperref}

\usepackage{graphicx}% Include figure files
\usepackage{dcolumn}% Align table columns on decimal point
\usepackage{bm}% bold math

\begin{document}

\preprint{APS/123-QED}

\title{Emergent tri-criticality in magnetic metamaterials}
%interplay of internal and external degrees of freedom}
%Phase transitions in magnetic metamaterials: Interplay of internal and external degrees of freedom}%\\ or \\ Metamaterials with Vanishing Spin Magnetic Moments}
%\title{Interplay of Internal and External Degrees of Freedom in Magnetic Metamaterials}%

\author{Bj\"orn Erik Skovdal}
\email{bjorn\_erik.skovdal@physics.uu.se}
\affiliation{Department of Physics and Astronomy, Uppsala University, Box 516, 751 20 Uppsala, Sweden}

\author{Gunnar K. Pálsson}
\affiliation{Department of Physics and Astronomy, Uppsala University, Box 516, 751 20 Uppsala, Sweden}

\author{P.C.W. Holdsworth}
\affiliation{Universit\'e de Lyon, ENS de Lyon, CNRS, Laboratoire de Physique, 69342 Lyon, France}

\author{Bj\"{o}rgvin Hj\"{o}rvarsson}
\affiliation{Department of Physics and Astronomy, Uppsala University, Box 516, 751 20 Uppsala, Sweden}

\begin{abstract}
Metallic discs engineered on the $100$ nm scale have an internal magnetic texture which varies from a fully magnetized state to a vortex state with zero moment. The interplay between this internal structure and the inter-disc interactions is studied in magnetic metamaterials made of square arrays of the magnetic discs. 
The texture is modeled by a mesospin of varying length with $O(2)$ symmetry and the inter-disc interaction by a nearest neighbour coupling between mesospins. The thermodynamic properties of the model are studied numerically and an ordering transition is found which varies from Kosterlitz-Thouless to first order via an apparent tri-critical point. The effective critical exponent characterising the finite size magnetization evolves from the value for the 2D-XY model to less than half this value at the tri-critical point. The consequences for future experiments both in and out of equilibrium are discussed. 
\end{abstract}

\date{\today}

\maketitle

\section{Introduction}
Universality, phase transitions and emergent magnetic properties are examples of phenomena that have recently been explored in metamaterials \cite{kapaklis_melting_2012,sendetskyi_continuous_2019,ostman_ising-like_2018,Arnalds_XY,Nisoli_2013,Nisoli:2017hg,ewerlin_magnetic_2013,leo_collective_2018,arnalds2016new,streubel_spatial_2018}. The ability to choose and investigate the effect of a single parameter, such as spin or spatial dimensionality\cite{sendetskyi_continuous_2019,ostman_ising-like_2018,Arnalds_XY,arnalds2016new}, as well as the possibility to directly observe individual magnetic elements has been a major impetus in this context \cite{wang_artificial_2006,farhan2013exploring,ladak_direct_2010,farhan2013direct,kapaklis2014thermal,morgan2011thermal,qi2008direct,morgan2013real}. These are indicators that nano-engineered materials can, in analogy with cold atom systems, become simulators of model many body problems, offering clear advantages over traditional condensed matter systems operating on the atomic scale. In this regard, metamaterials made up of magnetostatically interacting mesoscale islands, or \textit{mesospins}, are highly attractive. The multi-scale nature of the experimental set up allows for the emergence of new degrees of freedom from the internal spin textures, giving rise to rich behavior beyond that of standard magnetic models \cite{cowburn_single-domain_1999,sloetjes2021effect,Gliga_PRB_2015,shinjo_magnetic_2000,Klaui_vortx_2003,ding_magnetic_2005,ostman_hysteresis-free_2014}. 
Mesoscopic arrays of circular magnetic islands show a vast ensemble of internal magnetic textures, which vary strongly with the local environment. One of the most characteristic textures is a vortex which can progressively unwind from a state with zero magnetic moment into a collinear state with maximal moment \cite{skovdal2021temperature}.

The change from vortex to collinear states was shown to be driven by a competition between inter and intra-island interactions so that in an emergent description the interactions between mesospins are self-consistently modified by the collective environment \cite{skovdal2021temperature}.  As a consequence, mesospin ordering occurs via an emergent transition that depends on interactions at both the meso and atomic scales. In the initial experiments the transition was shown to be kinetic in nature, although a route towards true thermodynamic phase transitions was also identified. This suggests that the interplay between collective and internal energy scales could indeed open the door to phases and phase transitions that are not at present obtainable in atomic systems. 

In this paper we present a simple model which captures the essence of the interplay between the meso and the atomic length scales. The inter-island interactions are allowed to influence the net moment of the elements, which provides the coupling between the length scales involved. This leaves an XY spin model with an internal degree of freedom; the spin length, which can vary with an associated  energy scale. We find that, as a function of this internal energy scale, the magnetic phase transition evolves abruptly from Kosterlitz-Thouless (KT) to $1^\mathrm{st}$ order at a point showing a remarkable resemblance to a tri-critical point. That is, despite the absence of true long range magnetic order and the continuously varying spin length, the phase diagram closely resembles that of the $S=1$ Blume-Capel model \cite{Blume_PhysRev.141.517,Capel_1966966}. In this case, we observe an effective critical exponent, relevant for finite size systems, that varies continuously from that observed in 2D XY magnets of finite size, $\beta \approx 0.23$, towards a value characteristic of a tri-critical point.

\begin{figure}[t!]
\begin{center}
\includegraphics[width=1\linewidth]{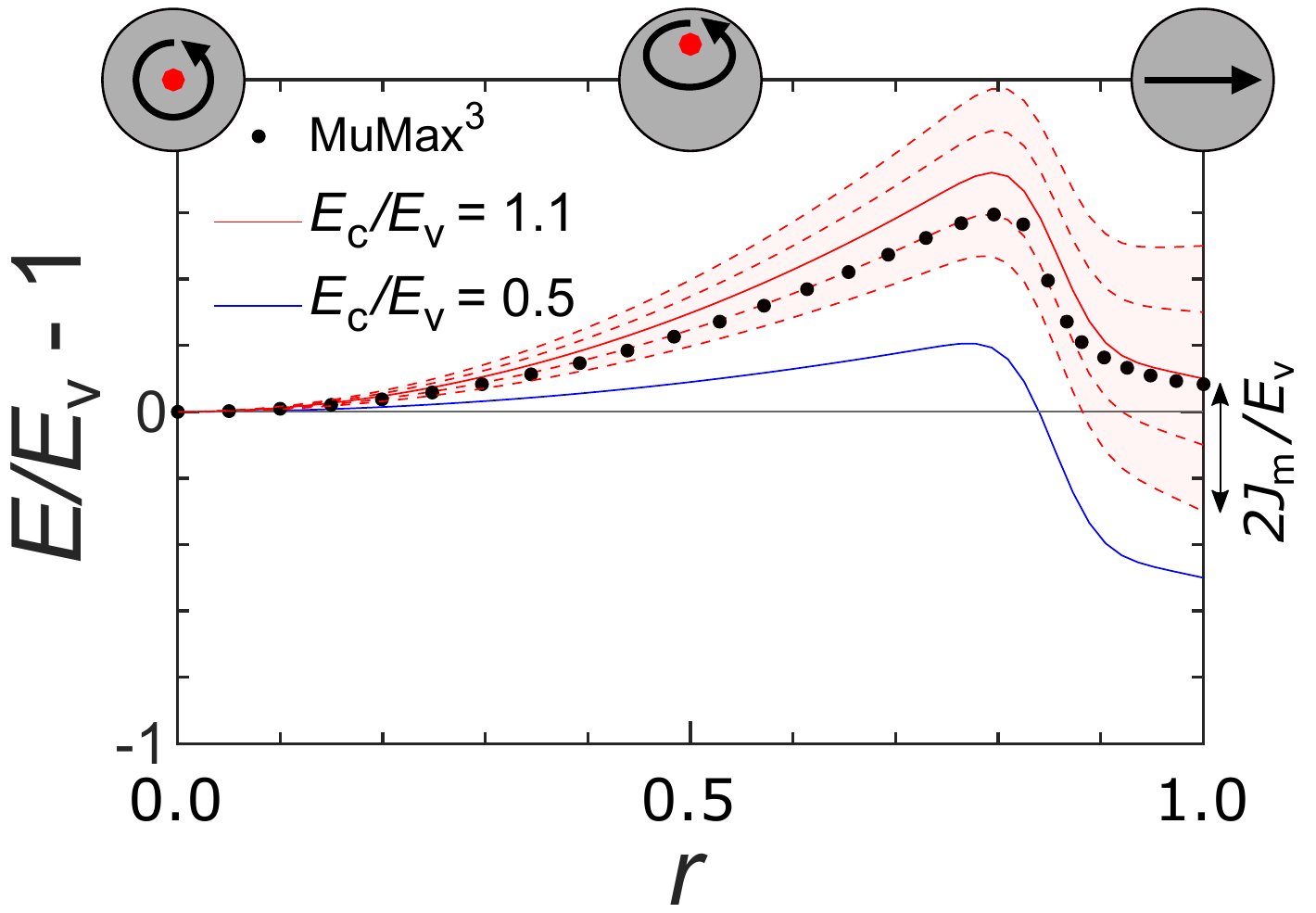}
%Loops_and_MvsT_2_ticks.pdf}
\caption{The black dots represent results from micromagnetic simulations for an isolated island with $E_\mathrm{c}/E_\mathrm{v}~\approx~1.1$. The solid (colored) lines represent the energy obtained from $S$ (Eq. \ref{lseq}) for an isolated island for two different values of $E_\mathrm{c}/E_\mathrm{v}$.  The red dashed lines represents the influence of parallel or antiparallel collinear neighbours for $E_\mathrm{c}/E_\mathrm{v}=1.1$.} 
\label{ls}
\end{center}
\end{figure}

\section{Modeling the magnetic metamaterial}

The total magnetic moment of interacting circular islands depends on the intrinsic material properties: internal magnetic texture, geometry, size and separation. For instance, above the inherent ordering temperature of the material there is no magnetic order at any length scale, while below that temperature, both thermal fluctuations and magnetic texture on the meso-scale are essential elements dictating the moment of the islands \cite{skovdal2021temperature}.
As examples a vortex state is a magnetic texture with a zero net in-plane moment, while a collinear inner state of the islands yields the largest net moment. The single vortex state can be characterized by two observables: a continuous change of the moment within the disc and a shift of the position of the vortex core, as illustrated in the top schematic of Fig. \ref{ls}. 
In the vortex state with a zero net moment, the vortex core sits at the center of the disc. The moment on the disc increases from zero as the vortex unwinds and approaches the disc edge. It is eventually annihilated as it moves across the edge of the magnetic island \cite{Tchernyshyov:2005gs,skovdal2021temperature,cowburn_single-domain_1999,ding_magnetic_2005,ostman_hysteresis-free_2014,shinjo_magnetic_2000}. 

In this paper we retain the variable moment length as the main manifestation of the evolving  magnetic texture, leaving the effects of the evolution in the vorticity for future work. The total moment on disc $i$ then becomes an in-plane vector $\vec M_{i}$, allowing for the definition of an in-plane, dimensionless mesospin vector of length
\begin{equation}
    r_i= \frac{|\vec M_i|}{M_\mathrm{max}},
\end{equation}
where $M_\mathrm{max}$ is the magnitude of the saturated total moment. The islands can thus be viewed as mesospins whose variable length depends on the internal spin texture of the discs. It varies continuously between zero and one, depending on the competition between internal and many body energy scales and a suitable model must include both these features.

The energy scale, $E$, associated with the variation of $r$ of an isolated disc has been studied in detail in previous work \cite{skovdal2021temperature}. 
The internal magnetic energy landscape of a single disc, obtained from micromagnetic simulations using MuMax$^3$ \cite{vansteenkiste_design_2014} is represented by the black dots in Fig. 1. Here we plot $E/E_\mathrm{v}-1$ {\it vs.} $r$, where $E_\mathrm{v}$ is the energy of the pure vortex state ($r=0$) and $E_\mathrm{c}$ is the energy of the collinear state ($r=1$). The energy landscape is highly asymmetric, with a maximum at approximately $r = 0.8$, corresponding to a vortex core positioned inside but close to the edge of the island. As the vortex core reaches the edge and moves outside the disc the energy associated with the magnetic texture rapidly decreases so that $E_\mathrm{c}$ lies well below the maximum energy. The ratio, $E_\mathrm{c}/E_\mathrm{v}$, determines if an isolated disc carries a moment or not in its lowest energy configuration and this can be varied either side of unity by changing the disc radius \cite{skovdal2021temperature}.

As shown in the figure, the landscape is qualitatively reproduced by the following phenomenological function 

\begin{equation}\label{lseq}
     S = E_\mathrm{c} \left(p\left(\frac{r}{r_0}\right)^2 - p + 1 \right) + \frac{E_\mathrm{v}}{3} \left(2p - r^2 + 1 \right),
     %S = (E_\mathrm{c}+1) \left(p\left(\frac{r}{r_0}\right)^2 - p + 1 \right) + \frac{(E_\mathrm{v}+1)}{3} \left(2p - r^2 + 1 \right) - 1,
     %S = (E_\mathrm{c}+1) \left(p\left(\frac{r}{r_0}\right)^2 - p + 1 \right) + \frac{1}{3} \left(2p - r^2 + 1 \right) - 1,
\end{equation}
where 
\begin{equation*}
    p = \frac{1}{2}\left(1 - \mathrm{erf}\left( \frac{r-r_0}{a\sqrt2} \right) \right),
\end{equation*}
with $a = 0.035$ and $r_0 = 0.85$.

The agreement between the phenomenological function, $S/E_\mathrm{v}-1$ (solid lines) and the micromagnetic simulations (black dots) is found to be good, as seen in the figure. Also included in the figure are the effects of inter-disc interactions (red dashed lines). Here the lines represent interactions of a disc with spin length $r$ with four fully collinear neighbors ($r=1$). Reversal of each neighbor from a parallel to an antiparallel state is therefore characterized by hopping from one line to the next in ascending order. The difference in energy is a measure of the many body interactions that one can expect in an array of discs. Of particular interest here is the case of $E_\mathrm{c}/E_\mathrm{v}=1.1$, which for the isolated disc indicates preference for a vortex state. When including interactions, however, a collinear configuration is instead favoured with the mesospins lying parallel to each other. This precursor illustrates how inter-disc interactions can influence the collective behaviour of an array promising the emergence of rich many body behaviour.

Magnetostatic inter-disc interactions are typically anisotropic in nature. However, here we assume isotropic nearest neighbour interactions. The choice is doubly motivated: simplicity combined with the possibility of designing hybrid metamaterials with the ascribed properties. 
With this approach, the Hamiltonian describing interactions between discs, placed on a square lattice taking into account both internal texture and many body interactions can be written 
\begin{equation}
    {H} = -J_\mathrm{m} \sum_{\langle ij \rangle}r_ir_j \cos(\theta_i-\theta_j) + \sum_{i}S_i(r). \label{Hamiltonian}
\end{equation}
$J_\mathrm{m}$ is the interaction between nearest neighboring islands and $\theta$ is the in-plane orientation of the mesospin,  $0 \leq  \theta <  2\pi$. 
The first term is similar to the 2DXY model, but includes the varying spin length, $0\leq r \leq1$ and the second term is the parameterized $r$-dependence defined by Eq. (\ref{lseq}). The proposed model is similar to the vector Blume-Capel model (VBCM) \cite{Blume_PhysRev.141.517,Capel_1966966,maciolek2004phase,Berker1979} in which vector spins take discrete lengths ($r=\{0,1\}$). The modeling of the emergent mesospin interactions gives us, in addition the continuous variation of $r$ and the phenomenological energy function $S(r)$.

Anticipating the situation where, below the bulk ordering temperature, the array of interacting discs can be thermally equilibrated, or that the non-equilibrium dynamics can be well represented through an emergent effective temperature, we study the thermal properties of the proposed model. The magnetization, $M$, is divided up into  the mesospin density $R$ and and orientation density $\Theta$, defined

\begin{equation*}
    M = \frac{1}{L^2} \sqrt{ \left( \sum_{i}r_i \cos \theta_i \right)^2 + \left( \sum_{i}r_i \sin \theta_i \right)^2  },  
\end{equation*}

\begin{equation*}
    R = \frac{1}{L^2}  \sum_{i}r_i ,  
\end{equation*}

\begin{equation} \label{eq1}
    \Theta = \frac{1}{L^2} \sqrt{ \left( \sum_{i} \cos \theta_i \right)^2 + \left( \sum_{i} \sin \theta_i \right)^2  } . 
\end{equation}
The susceptibilities are defined by:
  $ \chi_\mathrm{M} = L^2 \frac{\langle M^2 \rangle - \langle M \rangle^2}{T},$
  $  \chi_\mathrm{R} = L^2 \frac{ \langle R \rangle^2 -  \langle R \rangle^2}{T}, $
and, 
   $ \chi_\Theta = L^2 \frac{ \langle \Theta \rangle^2 -  \langle \Theta \rangle^2}{T}. $
$\Theta$ and $\chi_\Theta$ correspond to the magnetization and the magnetic susceptibility of the conventional 2D XY model.   It is also convenient to define the parameter $\Delta E = (E_\mathrm{c} - E_\mathrm{v} - 2 J_\mathrm{m})/E_\mathrm{v}$ whose sign designates the preference for broken symmetry or zero spin length in the lowest energy configuration. Using this definition, and with $J_\mathrm{m} = 0.2 E_\mathrm{v}$, degeneracy of the internal energy of the discs is obtained when $\Delta E = -0.4$ which marks a point of major importance in this work.

\section{Methods}

Arrays of $L^2$ mesospins with $L = 32$ on a square lattice with periodic boundaries were simulated using the Metropolis algorithm. Each calculation was based upon 40~000 thermalisation full lattice sweeps prior to 400~000 measurement sweeps, to ensure thermal equilibration and statistically robust results \cite{fernandez1986critical,dillon2010monte}. A full lattice sweep entails attempting to update both $r$ and $\theta$ once for each lattice site. The required thermalisation time scales were established through monitoring the relaxation time scale for $M$. $E_\mathrm{c}/E_\mathrm{v}$ was varied while always keeping $J_\mathrm{m} = 0.2 E_\mathrm{v}$. The initial state at each temperature was set to either a random spin configuration with respect to both $\theta$ and $r$ in a ``hot start'', or an ordered, fully magnetized  configuration in a ``cold start''.

\section{Results}

\begin{figure}[t!]
\begin{center}
\includegraphics[width=1\linewidth]{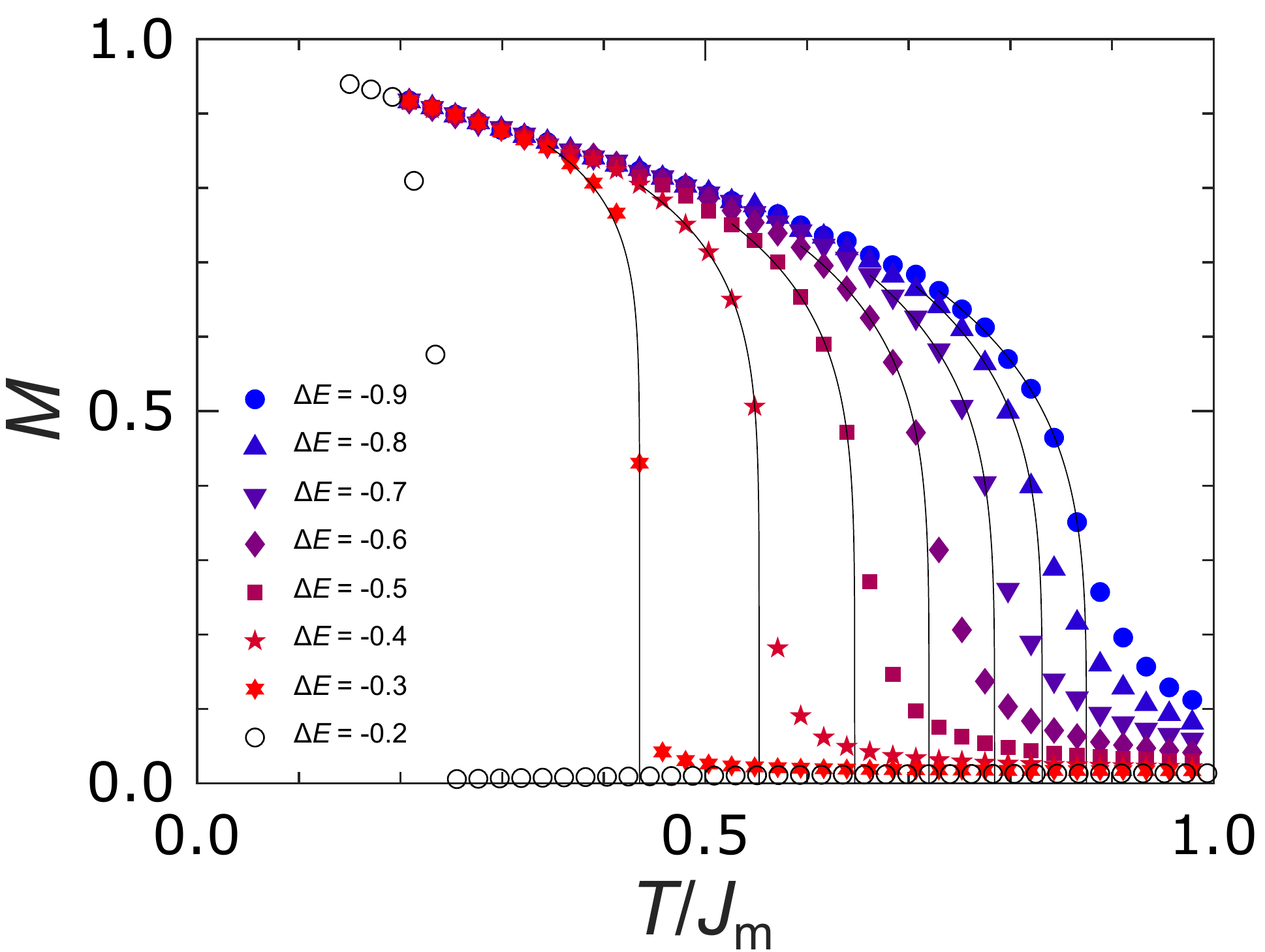}
\caption{Transitions for different values of $\Delta E$. The transition becomes sharper with increasing $\Delta E$, and becoming discontinuous for $\Delta E = -0.2$.} 
\label{beta}
\end{center}
\end{figure}

The magnetization, $M$, obtained from simulation is shown in Fig. \ref{beta} for different $\Delta E$ values from a hot start. The results reveal transitions from a high temperature disordered phase to a low temperature quasi-ordered phase. For negative values of $\Delta E$, a ferromagnetic ground state is energetically favourable. For $\Delta E$ large and negative the transitions are smooth, with the finite size magnetization resembling that observed through the KT transition of the 2D XY model, or plane rotator model \cite{bramwell1993magnetization,bramwell1994magnetization}. Increasing $\Delta E$ causes a sharpening of the transition, up to an apparent tri-critical point with $\Delta E \approx -0.3$. Increasing beyond this value, the finite size magnetization undergoes a discontinuous jump, as in a first order transition, see for example $\Delta E = -0.2$. For higher values ($\Delta E \geq -0.1$), the transition into an ordered collinear phase does not occur. 

\begin{figure}[t!]
\begin{center}
\includegraphics[width=0.6\linewidth]{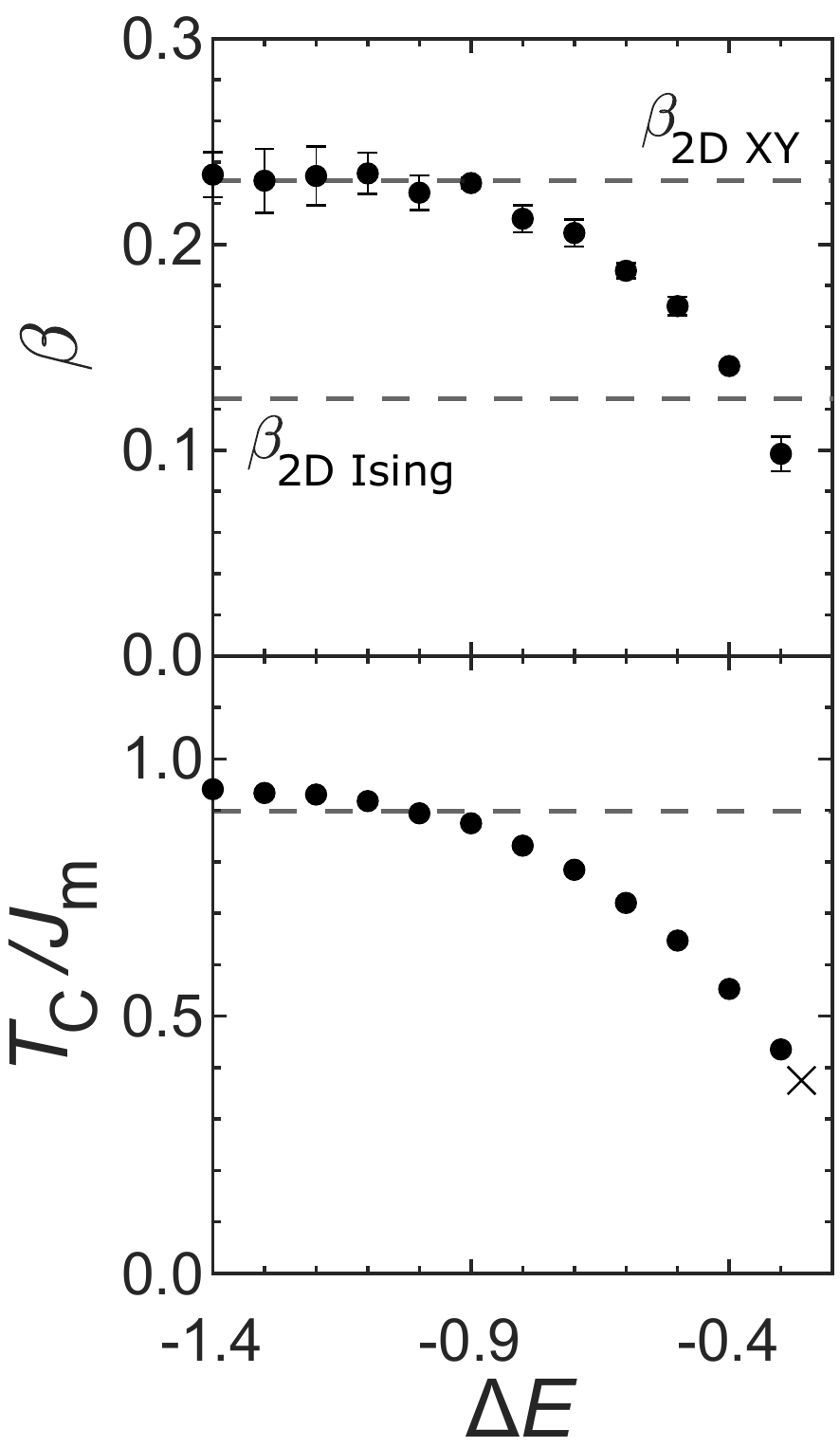}
\caption{Top panel: The effective critical exponent $\beta$ as a function of $\Delta E$.  The upper dashed line shows $\beta= 0.231$, the effective exponent of the 2D XY model.  The lower dashed line shows $\beta=1/8$, the value for the 2D Ising model. Bottom panel: The finite size ordering temperature, $T_\mathrm{C}$. The dashed line shows  $T_{KT}=0.898 J_\mathrm{m}$, the extrapolated value for the 2D XY model. The cross marks the estimated tri-critical temperature. }
\label{insets}
\end{center}
\end{figure}

The curves in Fig. \ref{beta} are fits to the data sets of the form $M=M_0(T-T_\mathrm{c})^\beta$, where $T_\mathrm{c}$ and $\beta$ are free parameters and $M_0=1$. Given this phenomenology, one should perhaps consider these curves as guides to the eye, although for $\Delta E$ large and negative the results are consistent with the zero parameter fitting procedure outlined in ref.[\onlinecite{bramwell1993magnetization}] as well as with many experimental observations \cite{Taroni_2008}. However, rather provocatively, the observed effective exponent $\beta$ does evolve in a way that is perfectly compatible with observations in a finite system as it crosses over from critical to tri-critical behaviour. The best fit exponents for a few values of $\Delta E$ are shown in the upper panel of  Fig. \ref{insets}. Starting from the expected value for the finite 2D XY model for $\Delta E <-0.9$, the fitted $\beta$ decreases continuously to less than half its initial value, with $\beta \approx 0.1$ close to the apparent tri-critical point. This evolution should be compared with mean field theory where the tri-critical exponent, $\beta_\mathrm{tri}=1/4$, down from $\beta=1/2$ at the regular critical point and with the 2D BCM where $\beta_\mathrm{tri}=1/24$ \cite{Ejima2018} is only one third of the 2D Ising critical exponent $\beta=1/8$. We note that the effective tri-critical exponent is quite close to the critical exponent for the Ising model, $\beta=1/8$, although it is difficult to incorporate this observation into a tri-critical scenario for quasi-ordering of the rotors.

In the lower panel of Fig. \ref{insets} we show the evolution of the transition temperature as $\Delta E$ increases. For $\Delta E<-0.9$, $T_\mathrm{C}$ slightly overshoots the extrapolated transition temperature value for the plane rotator model $T_\mathrm{C} \approx 0.898 J_\mathrm{m}$ \cite{Gupta96}. The overshoot is consistent with the expected logarithmic shift in the effective transition temperature with system size \cite{bramwell1993magnetization}. $T_\mathrm{C}$ then decreases with increasing $\Delta E$ reaching approximately half the plane rotator transition temperature at the tri-critical point before dropping discontinuously to zero for $\Delta E$ between $-0.3$ and $-0.2$. 

\begin{figure}[t!]
\begin{center}
\includegraphics[width=1\linewidth]{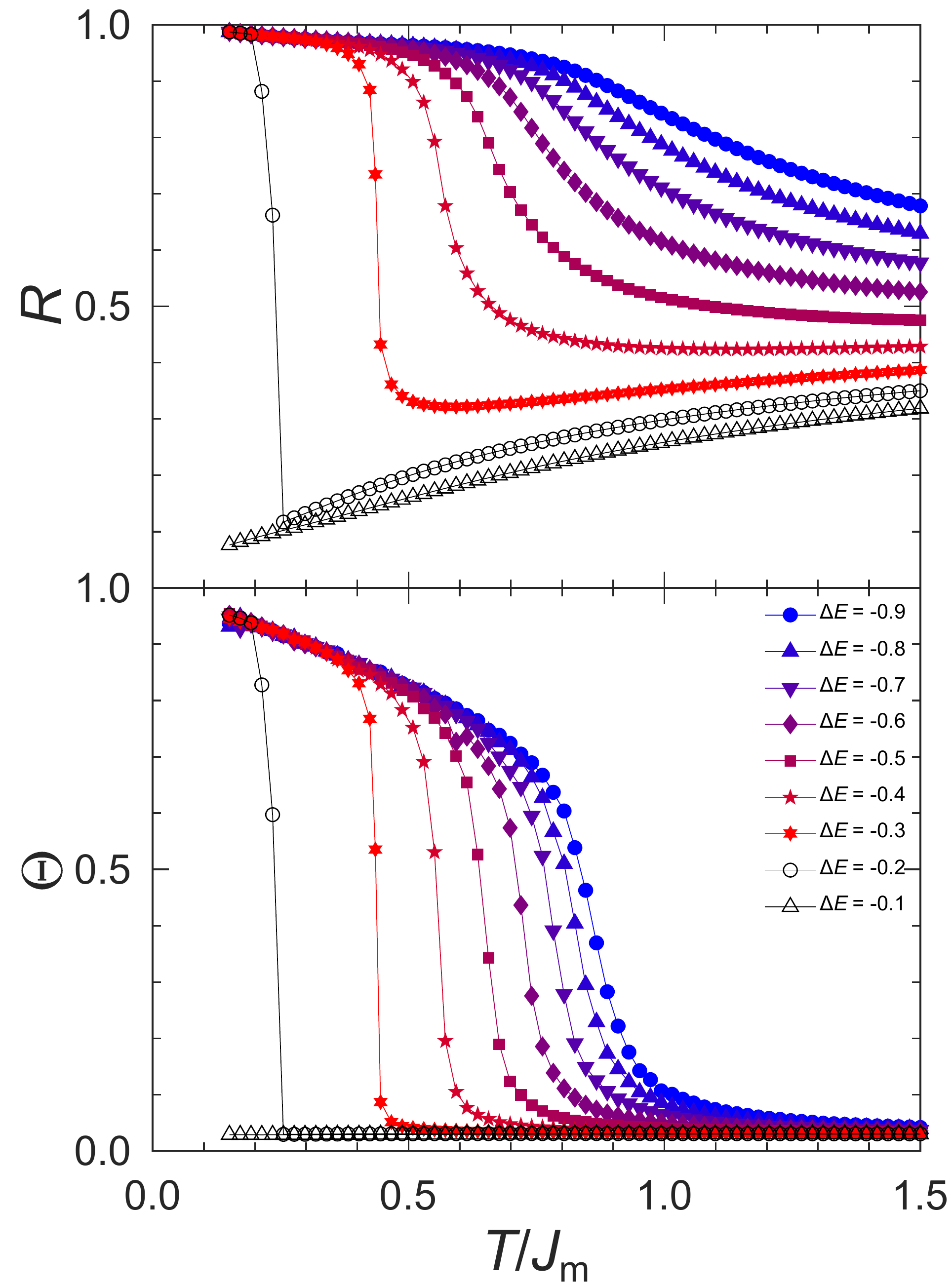}
\caption{Top panel: The average mesospin length $R$ {\it vs.} temperature for different $\Delta E$ . Bottom panel: The orientation density $\Theta$ {\it vs.} temperature for different $\Delta E$. The open symbols indicate values of $\Delta E$ in the $1^{st}$ order region of the phase diagram.}
\label{r_th}
\end{center}
\end{figure}

The change in the nature of the transition from KT to $1^{st}$ order is driven by the change in rotor length, as shown in the upper panel of Fig. ~\ref{r_th}, where we plot $R$ {\it vs} $T$ for different $\Delta E$ values. As $\Delta E$ passes through $-2 J_\mathrm{m} = -0.4$ the role played in the free energy by the rotor length changes. For $\Delta E < -0.4$, placing a rotor of maximum length leads to a gain in internal energy for both random and correlated spin configurations. As a consequence, the internal energy favours an ordered state with $R=1$, while entropic forces drive $R$ below unity, with maximum entropy for $R=0.5$. For $\Delta E$ considerably greater than -0.4, energy costs are such that $R\rightarrow 0$ as $T \rightarrow 0$ so that entropy drives the growth in $R$ at all finite temperatures. Between these two limits there is a small window of $\Delta E$ for which finite $R$ is energetically favorable {\it if} symmetry is broken (or almost broken in the case of a KT transition), otherwise it is entropically driven and energetically unfavourable. This phenomenology is well illustrated in Fig. ~\ref{r_th} which shows $R$ to be a monotonically increasing function as temperature is reduced for $\Delta E<-0.6$. For greater values of $\Delta E$, $R$ dips below $0.5$ and for $\Delta E=-0.3$, the approximate tri-critical value, $R$ clearly decreases as $T$ falls to intermediate values before rebounding to large values through the phase transition. 
It is this ``elastic'' resistance to large $R$ values at low temperature that drives the transition first order. At $\Delta E=-0.2$ the transition is clearly $1^{st}$ order, while for $\Delta E=-0.1$ no symmetry breaking is observed and $R$ decreases monotonically to zero as $T\rightarrow 0$.

This phenomenology in which energy gain at large $R$ depends explicitly on symmetry breaking, is generic to tri-critical systems. Similar behaviour is observed for the VBCM in two-dimensions (see Appendix) and in systems with discrete symmetry \cite{Brooks2014}. However, there are effects specific to the model studied here which allows for continuous variation of the rotor length. Close examination of Fig.~\ref{r_th} shows that $R$ drops below $0.5$ at intermediate temperature, even for $\Delta E=-0.5$ and $\Delta E=-0.4$, while in the equivalent figure for the VBCM (see Appendix), $R$ remains greater than $0.5$ for all $\Delta E\leq-0.4$. This difference arises as the system profits from the entropy associated with a continuous spread of rotor lengths, despite the energy gain from placing rotors of fixed length $r=1$. This is clearly a non-universal effect depending on the form of $S(r)$ and could be modified in different nano-engineered arrays. 

The KT transition is, from a thermodynamic point of view extremely special as there is no true magnetic symmetry breaking and so no order parameter in the thermodynamic sense. At first sight this might suggest that a $1^{st}$ order transition signaled by a discontinuous jump in such a parameter should be excluded. However, the day is saved here by the parameter $R$, which is a scalar measure of the mesospin density and which is a well defined intensive thermodynamic variable at all temperatures. A thermodynamic signal of the first order transition is therefore a discontinuous jump in $R$. However, the spin density remains coupled to the magnetization and to the development of quasi-long range orientational order for the mesospins through the tri-critical point and into the first order regime. This is illustrated in the lower panel of Fig. ~\ref{r_th}, where we show the evolution of $\Theta$ with temperature for different values of $\Delta E$. The purely orientation order parameter mimics $M$, with the pseudo-critical range narrowing as the tri-critical point is approached.

\begin{figure}[t!]
\begin{center}
\includegraphics[width=1\linewidth]{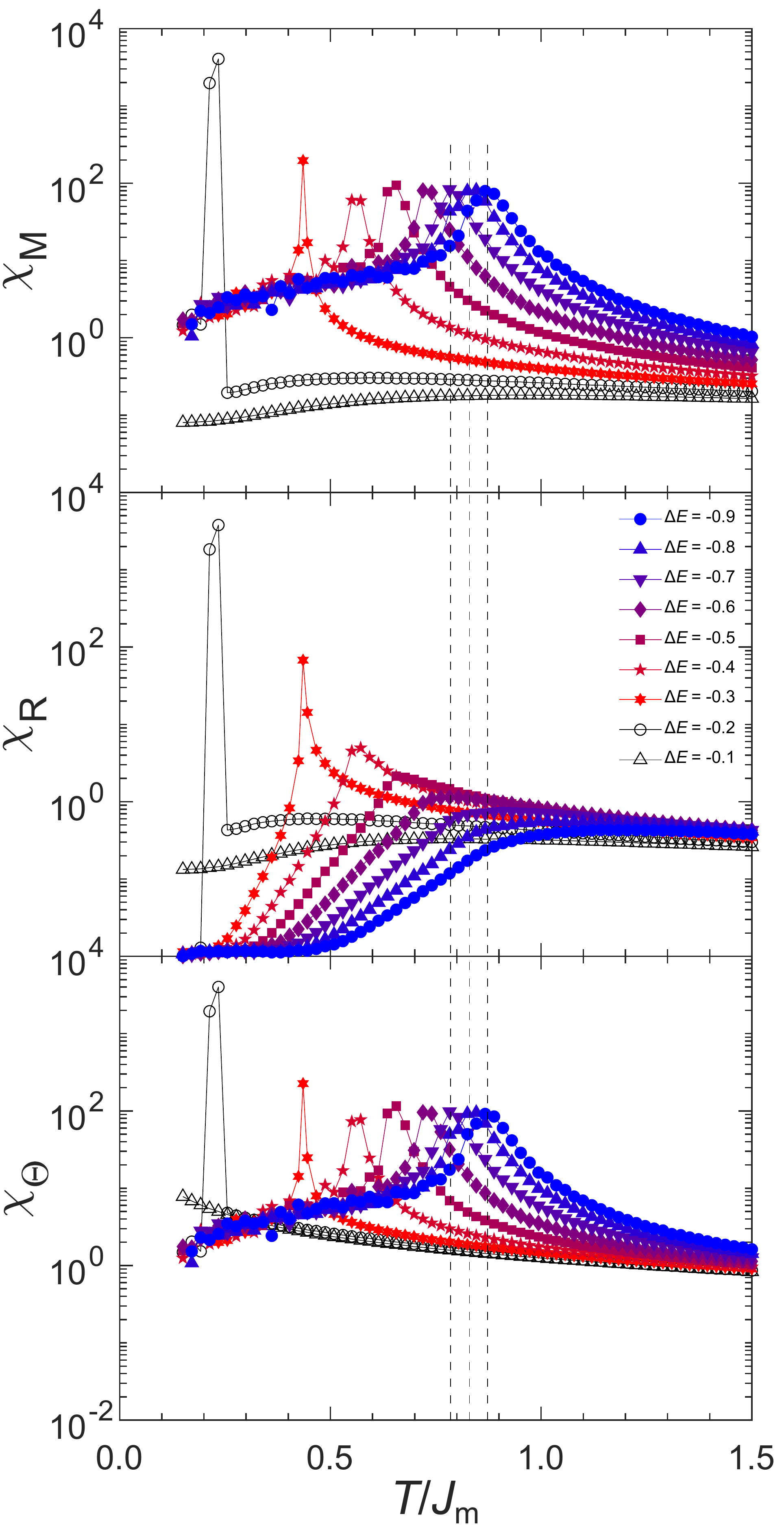}
\caption{Susceptibility {\it vs.} temperature for different $\Delta E$.Top panel: The magnetic susceptibility, $ \chi_\mathrm{M}$. Middle panel: The mesospin density susceptibility $  \chi_\mathrm{R}$. Bottom panel: The orientation density susceptibility  $\chi_\Theta$. The vertical dotted lines show the peak positions overlapping in $\chi_\mathrm{M}$ and $\chi_\Theta$, but not in $\chi_\mathrm{R}$ for the three lowest value of $\Delta E$. }
\label{sus}
\end{center}
\end{figure}

The development of tri-critical coupling between spin density and spin rotation degrees of freedom is illustrated by the three susceptibilities shown in Fig. ~\ref{sus}. For $\Delta E<-0.6$, which is deep in the KBT region, $\chi_\mathrm{M}$ and $\chi_\mathrm{\Theta}$ show a finite size rounded divergence at the same temperature, which can be fitted to the characteristic exponential form for the KT transition (not shown), while $\chi_\mathrm{R}$ shows a rounded maximum at temperatures that are decoupled from the KT transition and two orders of magnitude smaller than the singular susceptibilities. As $\Delta E$ increases into the crossover region towards tri-criticality, a sharp peak in $\chi_\mathrm{R}$ emerges. It rapidly locks onto the divergences in $\chi_\mathrm{M}$ and $\chi_\mathrm{\Theta}$ which also sharpen so that, on arriving at the apparent tri-critical point the three susceptibilities show the same sharply singular feature. This can be taken as a signature of the coupling of the internal and external degrees of freedom of the mesospins. 

\begin{figure}[t!]
\begin{center}
\includegraphics[width=1\linewidth]{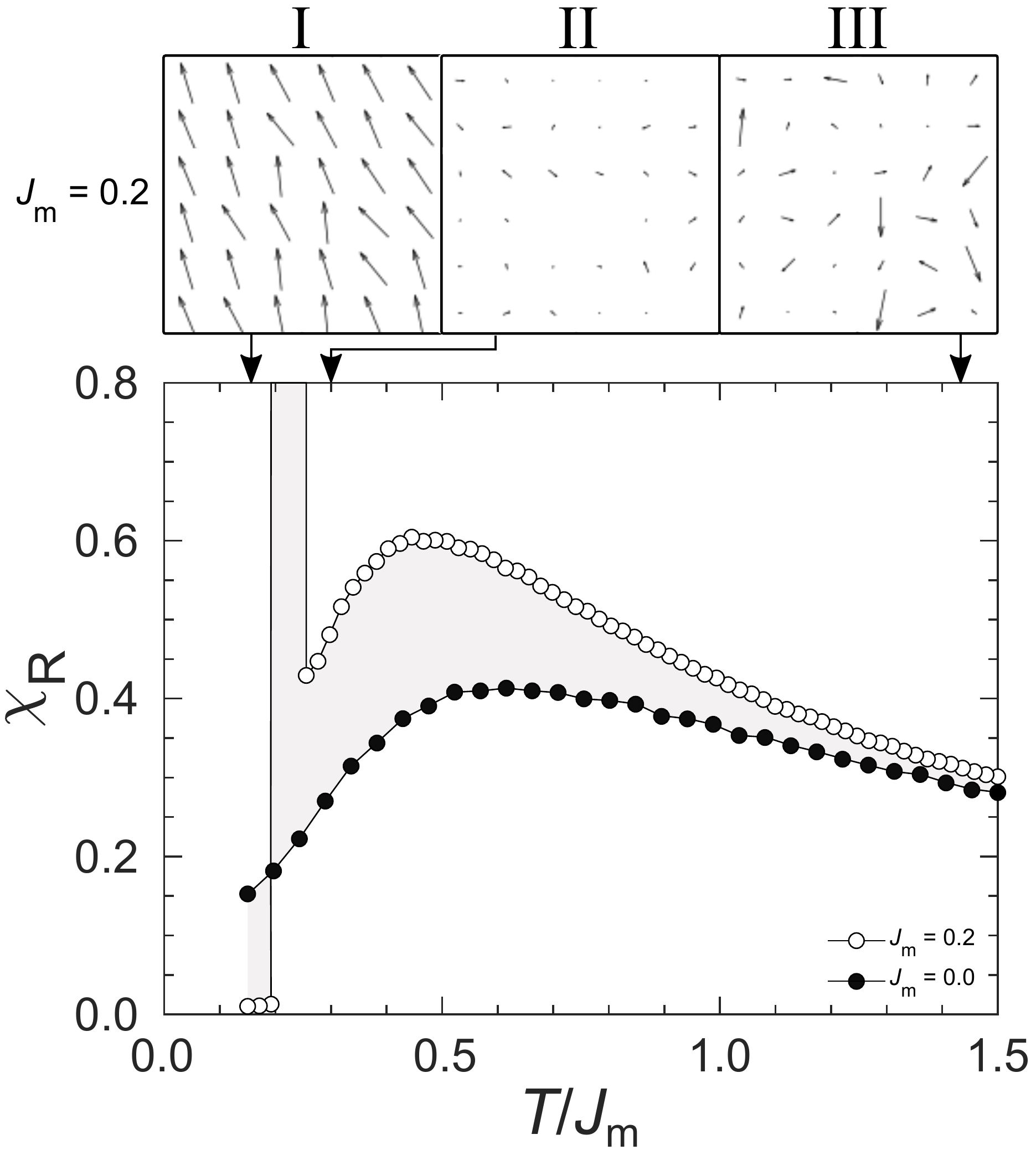}
\caption{Upper Panel: snapshots of mesospin configurations at high (III), intermediate (II) and low temperatures (I). In (III) a full range of $r$ values are observed, while in (II) the range is limited to small values. Transformation from (III) to (II) is a crossover. Evolution from (II) to (I) is via a $1^{st}$ order transition for $J_m±
\ne 0$. Lower Panel: $\chi_\mathrm{R}$ as a function of temperature for $\Delta E = -0.2$, with and without interactions.}
\label{3phase}
\end{center}
\end{figure}

Within the first order regime, the three regions; the high temperature entropic regime, the unfavorable intermediate temperature regime and the broken symmetry ordered phase are illustrated in Fig.~\ref{3phase} in the upper panel. The figure shows snapshots for different temperatures, for $\Delta E = -0.2$. As we are in the regime where, in the absence of interactions, the vortex state is favoured, the low energy magnetic state is generated through many body interactions. In region III both orientational disorder and a wide range of rotor lengths can be observed. In region II while the rotors remain disordered their mean length is clearly reduced, reflecting the energy cost of creating full length rotors while remaining in the disordered phase. In the low temperature phase, I, the symmetry breaking allows for an energy gain on generating extended rotors. In the lower panel we show $\chi_\mathrm{R}$ on a linear scale over the same temperature range. Entry into the intermediate range is marked by a broad maximum at around $T=0.5 J_\mathrm{m}$ signaling a rapid reduction in mean rotor length. Below the peak,  $\chi_\mathrm{R}$ decreases until it hits the first order discontinuity at $T\approx 0.25 J_\mathrm{m}$ below which the rotor length remains more or less fixed near the maximum value. 
Also shown is the rotor susceptibility for a non-interacting system ($J_\mathrm{m}=0$). The peak at intermediate temperature is lower and broader when interactions are switched off, illustrating that rotor-rotor interactions offset the energy cost of finite spin length, helping to maintain their presence down to lower temperatures.

\begin{figure}[t!]
\begin{center}
\includegraphics[width=1\linewidth]{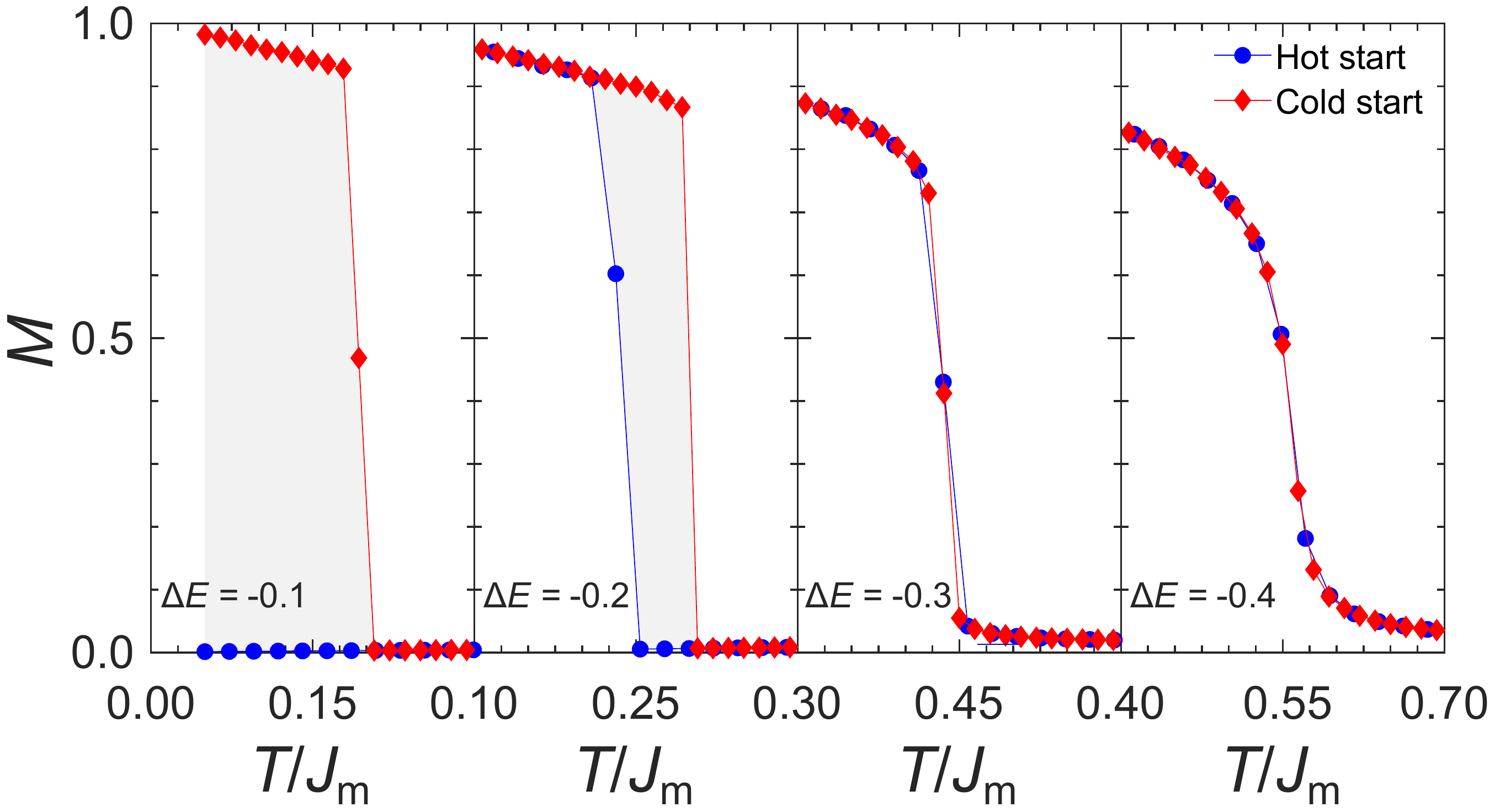}
\caption{$M$ as a function of temperature for different values of $\Delta E$, starting at random (hot start) or collinear (cold start) spin configurations at each temperature. }
\label{hotncold}
\end{center}
\end{figure}

The data shown in Figs. \ref{beta}, ~\ref{r_th} and ~\ref{sus} is for hot starts for fixed $\Delta E$. Following this protocol, the symmetry breaking transition disappears between $\Delta E=-0.2$ and $-0.1$ even though the ground state of an ordered configuration with rotors of unit length remains lower than any disordered state for $\Delta E < 0$. The loss of the transition in this range is a non-equilibrium result characteristic of a first order transition and the presence of metastable states. Making runs from cold starts in the ordered configuration exposes hysteresis in $M$, due to the loss of ergodicity as shown in Fig. \ref{hotncold}. For $\Delta E = -0.1$ the ordered state survives a cold start up to $T=0.15 J_\mathrm{m}$ and the metastability survives up to around $\Delta E = -0.3$ in the tri-critical region.

\begin{figure}[t!]
\begin{center}
\includegraphics[width=1\linewidth]{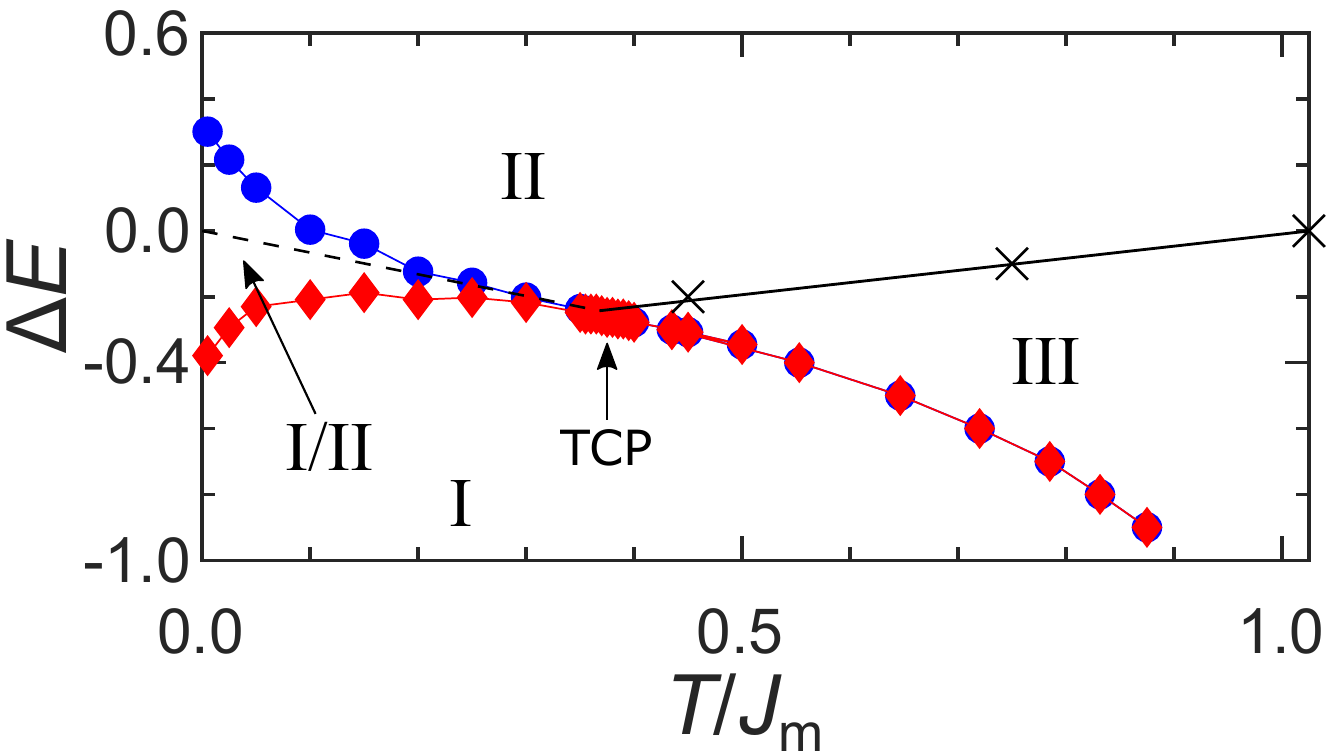}
\caption{The phase diagram showing the tricritical point (TCP) separating the first- and second-order behavior. The dashed line represents the phase boundary in region of $1^{st}$ order transitions and the area between blue (cold start) and red (hot start) points marks the region of metastability. The three X's indicate the positions of the broad maximum in $\chi_\mathrm{R}$ (shown in Fig. \ref{3phase} for $\Delta E = -0.2$) for $\Delta E = \{-0.2, -0.1, 0.0\}$. The solid line connecting the crosses is a guide to the eye. }
\label{pd}
\end{center}
\end{figure} 

The hysteresis can also be observed by ramping $\Delta E$ in loops from negative values upwards and back again, while holding the temperature fixed \cite{dillon2010monte}.
The full phase diagram in the $\Delta E$, $T$ plane from loops ramping $\Delta E$ at fixed temperature is shown in Fig. \ref{pd}. This phase diagram is similar to that observed for the VBCM \cite{dillon2010monte}. The ordered phase resists finite temperature fluctuations up to $\Delta E \approx 0.3$, considerably above the equilibrium threshold for stability of the ordered phase. The figure therefore shows a finite region of metastability in which the ordered and disordered phases, labeled I and II coexist for simulations of fixed time scale. The zone of metastability closes at the tri-critical point and ramping $\Delta E$ gives an alternative measure of its position, which we estimate to be $\Delta E = -0.26 \pm 0.01$, $T_{tri} = 0.375 \pm 0.025$. This is in good qualitative agreement with our estimate from the evolution of the effective critical exponent. The true phase boundary must run close to the line extrapolating between  $\Delta E=0$ at $T=0$ and the tri-critical point, as shown in Fig. \ref{pd}, although we have not attempted to evaluate it in detail. The crossover between regions II and III in which the rotors are confined to short lengths and in which a full spread of lengths appear is also shown. The position of the broad maximum in $\chi_\mathrm{R}$ which characterizes the crossover is marked for $\Delta E = \{-0.2, -0.1, 0.0\}$.

\section{Discussion}

We have shown that engineered two-dimensional arrays of magnetic discs on the mesoscale offer interactions that map convincingly onto a model system showing tri-criticality. In this development, we represent inter-disc interactions and internal spin textures by an effective nearest neighbour coupling between magnetic mesospins of varying length. The energy scale fixing the mesospin length is the magnetic vortex core energy, giving a Blume-Capel type model (BCM) \cite{Blume_PhysRev.141.517,Capel_1966966} with both continuous, in-plane rotor orientations \cite{dillon2010monte,Chamati2007} and continuous rotor length. The tri-criticality observed numerically is rather special in that it marks the evolution from a Berezinski-Kosterlitz-Thouless phase transition to a first order transition \cite{dillon2010monte,Chamati2007,Santos2018}. 

Tri-critical systems are the confluence of three phases whose thermodynamics is governed by three independent thermodynamic variables \cite{Griffiths1970}. As a consequence their critical properties are chracterised by three scaling variables \cite{Riedel72} and associated critical exponents \cite{Cardybook}. This situation is captured most simply by the Blume-Capel model \cite{Blume_PhysRev.141.517,Capel_1966966}, in which temperature and field conjugate to an order parameter with $Z_2$ symmetry are joined by an energy scale or chemical potential associated with spin creation and annihilation. 

Archtypical examples of  tri-crtical systems are the merger of the super-fluid transition of $^4$He and the critical point of the demixing transition in $^4$He-$^3$He mixtures \cite{Graf1967,Garcia2002},
or the smectic C$^{\ast}$ - smectic A transition, which evolves through tri-criticality on mixing two species of liquid crystal \cite{Shashidhar1988}.
Experimental studies of the tri-criticality are complicated by the difficulty in accessing the three intensive thermodynamic variables. For example for $^4$He-$^3$He mixtures, while exquisite temperature control is possible  \cite{Lipa1996}, the field conjugate to the superconducting order parameter is inaccessible. The mixture can be controlled by varying the $^3$He mole fraction which serves as a second order parameter, but the true intensive variable, the chemical potential difference, $\mu=\mu_{^3\mathrm{He}}-\mu_{^4\mathrm{He}}$ is also inaccessible. In principle, $\mu$ could be controlled in ultra-thin helium films \cite{Bishop1978} but the experimental environment is extremely challenging. In other systems, the situation is even more constrained. Arrays of Josephson junctions can be diluted \cite{Yun2006}, allowing the approach to tri-criticality, but the procedure introduces quenched site disorder and the complexity associated with it.  First order magnetic transitions occur, for example in FeRh films \cite{Maat2005} or in spin ice materials \cite{Spinicebook}. These transitions could be tuned to tri-criticality \cite{Borzi2016,Raban2019}, but this would require control of both the coupling constant and the chemical potential through applied pressure or site dilution. 
%Finally cold atom experiments offer the promise of the kind of control demanded for Boson and Fermion mixtures \cite{cold-atom-review2021} in future experiments. 

Our work opens the door to experiments in which all three intensive thermodynamic variables relative to the tri-critical phase diagram can be controlled through the change of disc spacing, radius and thickness \cite{skovdal2021temperature} and application of an external field. %In this regard, there is already some experimental evidence in line with the phase diagram we present here, where large discs display vortex states, small discs display collinear states, and intermediate sized discs display coexistance of both
A change of symmetry for rotor orientations from continuous to discrete and a return to the original BCM is also envisageable through changing the disc shape.

BCM and vector-BCM (VBCM) models have been further extended to Blume-Emergy-Griffiths (BEGM) type models \cite{Blume1971,Berker1979,Cardy1979,maciolek2004phase}
to describe the full $^4$He-$^3$He phase diagram. Here, a bi-quadratic interaction between spins is added capturing isotropic interactions. The extra term introduces the possibility of separating the demixing from the ordering of the internal degree of freedom, allowing for liquid-gas like criticality, $^4$He-$^3$He tri-criticality and a triple point between superfluid, normal $^4$He rich and $^3$He rich phases in both three \cite{maciolek2004phase} and in two dimensions \cite{Chamati2007,Santos2017,Santos2018}. The extra bi-quadratic term could also be engineered in the nano-arrays by modification of the disc topology, leading to the development of quadrupole interactions. This would allow for the experimental study of models resembling the BEGM and vector-BEGM in two-dimensions.

The magnetization of a system of finite size is a prime experimental indicator of the Kosterliz-Thouless phase transition  \cite{bramwell1993magnetization}. It changes in a characteristic manner though the transition, with the emergence of an effective magnetic critical exponent $\beta\approx0.23$ \cite{Taroni_2008}. We have shown here that the effective magnetization exponent crosses over towards tri-criticality in an analagous manner to a thermodynamic exponent, with an effective tri-critical value less than half the critical value, as is the case for the two-dimensional BCM \cite{Kwak2015,Ejima2018}. Going beyond this phenomenology would require more extensive numerics and a deeper examination of the theory. This paper provides a platform for this in future work, but more importantly for the present, this straightforward approach provides a platform for experiments on arrays of mesoscale discs in which order parameter crossover, effective or otherwise, should be accessible to measurement. 

Further, system size will be an independent control parameter of these metamagnetic systems, providing  experimental access to finite size scaling. This is
a powerful tool for simulation \cite{Kwak2015}, but is generally outside the realm of experiment in condensed matter systems. Artificial systems such as the mesospin arrays presented here or cold atom platforms \cite{cold-atom-review2021} could provide future access to this essential phenomenology.

\section{Conclusion}

The experimental realisation of emergent tri-criticality in magnetic metamaterials with coupled intra and inter-island excitations poses serious experimental challenges, notably the creation of an environment showing equilibrium thermodynamics (real or effective) and controlled departures from it. However it offers a test case that prepares the ground for a vast array of possibilities offered by nano-engineered metamagnets, with both  fundamental exploration and technological applications in mind. In particular, the identification of
mesospins of variable length in controlled out of equilibrium environments invites applications in
adaptive matter \cite{kaspar2021rise} through the dynamical modification of the many body energy landscape \cite{walther2020responsive}.
The self-modification of the energy discussed here is analogous to the interstitial self-trapping of hydrogen in metals, where the hydrogen interstitial and the local strain field form a dynamic quasi particle\cite{blomqvist2010significance}. Local energy landscape fluctuations within an array of mesospins could also offer local sensing capabilities and long term memory, a further 
cornerstones of adaptive or intelligent matter  \cite{kaspar2021rise}. Clever engineering of the metamaterials may therefore offer pathways towards more advanced materials or even analogue logic mesospin components\cite{bhanja2016non}.

\section*{Data availability}
All data are available from the corresponding author upon reasonable request.

\section*{Acknowledgments}
B.H. acknowledges support from the Swedish Research council (VR). P.C.W.H. is supported by the ``Agence Nationale de la Recherche'' under Grant No. ANR-19-CE30-0040. G.K.P gratefully acknowledges the Swedish Research Council (VR -  Vetenskapsrådet) grant number 2018-05200 and the Swedish Energy Agency grant 2020-005212 for funding.

\section*{Author contributions}
B.E.S conceived the model. B.E.S and G.K.P performed the simulations. P.C.W.H. provided additional context and interpretation of the results. B.H. supervised the project. B.E.S, P.C.W.H. and B.H. wrote the manuscript. All authors discussed the results and contributed to the manuscript.

\section*{Competing interests}
The authors declare no competing interests.

\appendix
\section{}
In the context of our work, the Hamiltonian for the VBCM is defined
\begin{equation}
    {H} = -J_\mathrm{m} \sum_{\langle ij \rangle}r_ir_j \cos(\theta_i-\theta_j) + \sum_{i}Dr_i. \label{Hamiltonian}
\end{equation}
with $D=E_\mathrm{c}-E_\mathrm{v}$. The spin length $r_i$ now takes on one of two values: $r_i=1$ or $r_i=\epsilon$,  {\it lim} $\epsilon \rightarrow 0$. Taking the limit $r_i \rightarrow 0$ attributes a phantom rotational degree of freedom to the vacancy,  ensuring that it occupies the same volume in configuration space as a site filled with a classical spin. With this precaution the integrals of the partition function can be normalized by $2\pi$ \cite{Berker1979}, so that $R\rightarrow 0.5$ at high temperature. As before, the dimensionless energy shift is defined $\Delta E=(D-2J_\mathrm{m})/E_\mathrm{v}$.

\begin{figure}[h!]
\begin{center}
\includegraphics[width=1\linewidth]{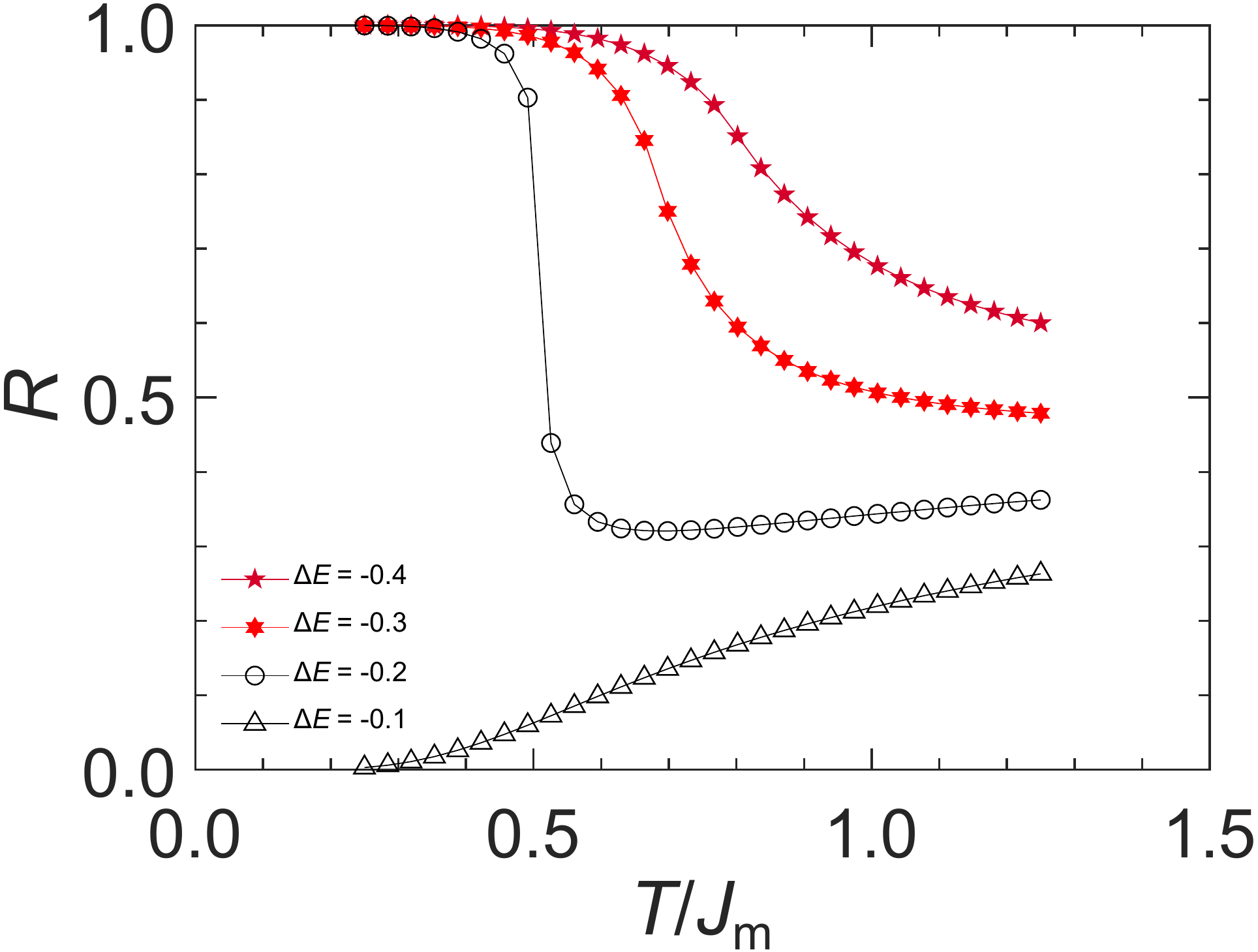}
%Loops_and_MvsT_2_ticks.pdf}
\caption{Results from the discretized model where $r=\{0,1\}$.}
\label{VBCM}
\end{center}
\end{figure}

The evolution of $R$ vs temperature for different $\Delta E$ values is shown in Fig. \ref{VBCM} confirming the predicted behaviour at high temperature. For $\Delta E \leq -0.4$, $R$ remains greater than $0.5$ over the whole temperature range. This is in contrast with  the model in the main text with continuous variation in $r_i$, in which $R$ dips below $0.5$ at intermediate temperatures for both  $\Delta E = -0.4$ and $\Delta E = -0.5$.

%\bibliographystyle{apsrev4-1}
%\bibliography{Mylib}


\end{document}